\documentclass[prl, reprint, amsmath,amssymb,nofootinbib,floatfix, aps, longbibliography]{revtex4-1}
\usepackage{graphicx}
\usepackage{dcolumn}
\usepackage{bm}
\usepackage[colorlinks=true,urlcolor=rossos,anchorcolor=black,citecolor=mygreen,filecolor=black,linkcolor=black,menucolor=blue,linktocpage=true]{hyperref} 
\usepackage{booktabs}
\usepackage{colortbl}
\usepackage{collcell}
\usepackage{enumerate}
\usepackage{float}
\usepackage{verbatim}
\usepackage{multirow}
\usepackage{xparse}
\usepackage{tabularx}

\definecolor{rossos}{cmyk}{0,1,1,0.55}
\definecolor{mygreen}{rgb}{0.27, 0.64, 0.48}
\definecolor{mygray}{gray}{.95}

\newcommand{\virg}[1]{``#1''}

\def\min{\text{min}}
\def\max{\text{max}}

\def\DM{\text{DM}}

\def\CM{\text{c.m.}}
\def\LOS{\text{LOS}}
\def\BH{\text{BH}}
\def\doppler{\mathcal{D}}

\begin{document}

\title{Direct Detection Constraints on Blazar-Boosted Dark Matter}

\author{Jin-Wei Wang}
\email{jinwei.wang@sissa.it}
\affiliation{Scuola Internazionale Superiore di Studi Avanzati (SISSA), via Bonomea 265, 34136 Trieste, Italy}
\affiliation{INFN, Sezione di Trieste, via Valerio 2, 34127 Trieste, Italy}
\affiliation{Institute for Fundamental Physics of the Universe (IFPU), via Beirut 2, 34151 Trieste, Italy.}

\author{Alessandro Granelli}
\email{agranell@sissa.it}
\affiliation{Scuola Internazionale Superiore di Studi Avanzati (SISSA), via Bonomea 265, 34136 Trieste, Italy}
\affiliation{INFN, Sezione di Trieste, via Valerio 2, 34127 Trieste, Italy}
\affiliation{Institute for Fundamental Physics of the Universe (IFPU), via Beirut 2, 34151 Trieste, Italy.}

\author{Piero Ullio}
\email{ullio@sissa.it}
\thanks{\\ J.-W.W. and A.G. contributed equally to this work.\\}
\affiliation{Scuola Internazionale Superiore di Studi Avanzati (SISSA), via Bonomea 265, 34136 Trieste, Italy}
\affiliation{INFN, Sezione di Trieste, via Valerio 2, 34127 Trieste, Italy}
\affiliation{Institute for Fundamental Physics of the Universe (IFPU), via Beirut 2, 34151 Trieste, Italy.}

\begin{abstract}

We explore the possibility that relativistic protons in the extremely powerful jets of blazars may boost via elastic collisions the dark matter particles in the surroundings of the source to high energies. We concentrate on two sample blazars, TXS 0506+056 — towards which IceCube recently reported evidence for a high-energy neutrino flux — and BL Lacertae, a representative nearby blazar. We find that the dark matter flux at Earth induced by these sources may be sizable, larger than the flux associated with the analogous process of dark matter boosted by galactic cosmic rays, and relevant to access direct detection for dark matter particle masses lighter than 1 GeV. From the null detection of a signal by XENON1T, MiniBooNE, and Borexino, we derive limits on dark matter-nucleus spin-independent and spin-dependent cross sections which, depending on the modelization of the source, improve on other currently available bounds for light dark matter candidates of 1 up to 5 orders of magnitude.

\end{abstract}
\maketitle

\noindent \textit{\textbf{Introduction.}}---
The nature of dark matter (DM) in the Universe remains elusive \cite{hep-ph/0404175, 1807.06209}. Steady progresses have been made in the attempt to identify the DM particles forming the Milky Way (MW) halo by detecting their elastic scattering of target nuclei, such as, most recently, by XENON1T \cite{1206.6288, XENON:2017lvq} and PandaX-II \cite{PhysRevLett.119.181302}. A limitation of such direct detection technique is the fact that MW DM particles are expected to have small velocities, typically $\sim 10^{-3} c$, and hence nuclear recoil energies exceed detector thresholds, say $\sim 1$~keV, only for DM masses $\gtrsim 1$~GeV. 

In the latest years a few scenarios with \virg{boosted} DM populations have been proposed, allowing for nuclear recoil signals even for lighter DM particles, see, e.g., Refs.~\cite{1405.7370,1506.04316, 1708.03642, 1709.06573}. In Ref.~\cite{PhysRevLett.122.171801} the authors considered the interesting possibility that MW DM particles are boosted via elastic scatterings with galactic high-energy cosmic rays, deriving relevant constraints for sub-GeV DM candidates. We propose here blazars as ideal DM boosters: they are associated with intense sources of high-energy nonthermal particles, they are located in a gravitational potential with a supermassive black hole (BH) at the center, whose formation may have triggered a large enhancement of the ambient DM density, and they are relatively close to us. 

Blazars are a type of active galactic nuclei (AGN) accelerating particles into two back-to-back jets, with one of them in close alignment to our line of sight (LOS) \cite{astro-ph/9506063}. They are characterized by a nonthermal continuous photon spectral energy distribution (SED) with two peaks, one in the infrared or x-ray bands and the other at $\gamma$-ray frequencies \cite{0912.2040}. Models of the SED~\cite{1992ApJ...397L...5M, 1993A&A...269...67M, 1996ApJ...461..657B, astro-ph/0004052, 0711.4112, 0909.0932, 1304.0605} have been refined with GeV-TeV data from Fermi-LAT and air cherenkhov telescopes~\cite{abdo_501,Abdo_2011}: it is widely accepted that the low-energy peak is due to synchrotron emission by electrons, but there is still no consensus on the origin of the high-energy component. While electrons could also be responsible for it (leptonic models), a highly relativistic population of protons may also be present in the jets and account for the $\gamma$-ray emission (pure hadronic and hybrid leptohadronic models, see, e.g., Ref.~\cite{2012.13302} for a recent model review).
Moreover, given the high variability of blazars (both in time and population), the parameters of each model, as well as the goodness of the fit, strongly depend on the considered source and the time of observation. It is therefore complicated to establish a unifying picture.

Fortunately, multimessenger astrophysics can provide more insights into the physics of blazar jets. For instance, in both hadronic and leptohadronic models energetic neutrinos can be produced through photo-meson production, while in purely leptonic models no neutrino appears. Therefore, the detection of neutrinos from a blazar is a smoking-gun signal for the presence of relativistic protons in the jet.
Recently, a very strong hint for the detection  
of high-energy cosmic neutrinos from the blazar TXS 0506+056 was found by the IceCube Neutrino Observatory \cite{IceCube1, IceCube2,1807.04461}. Studies of the SED have shown that the leptohadronic model
is in general adequate to explain both the detected neutrino flux and the $\gamma$-ray emission of TXS 0506+056~\cite{1807.04537, TXS, TXS_2, 1812.05939, 1908.10190, 1911.04010}. For these reasons, in this work we will concentrate on pure hadronic and/or leptohadronic models.

Electrons and protons in the jets of a blazar can collide with ambient DM particles. Scatterings off DM by electrons and protons in the jet plasma of AGN were already considered in Refs.~\cite{astro-ph/9706085, PhysRevD.82.083514}, where the authors focused on photon emissions. Instead, in the present Letter, similarly to the  acceleration mechanism due to cosmic rays \cite{PhysRevLett.122.171801}, we consider DM boosted by protons in the jet of blazars, derive the induced DM flux at Earth and compute the associated nuclear recoil direct detection signal. We refer to DM boosted via this mechanism as blazar-boosted dark matter (BBDM). Motivated by IceCube observations, we decide to focus our study on the blazar TXS 0506+056. For comparison, we also consider the near representative blazar BL Lacertae. We then discuss the implications the nondetection of BBDM from the two sources have on spin-independent and spin-dependent DM-nucleus cross sections. An analogous analysis dedicated to leptophilic DM, for which boosting by electrons in the jet and scattering off electrons in the detector are relevant, is postponed to a future related study \cite{Granelli:2022ysi}.\\

\noindent \textit{\textbf{Spectrum of the relativistic blazar jet.}}--- We consider the simplifying assumption that the blazar emission originates from a homogeneous zone (blob) in the jet where particles (mainly electrons and protons) are distributed isotropically  \cite{Dermer2009-DERHER}. 
The blob, as seen by an observer standing still with respect to the BH center of mass, propagates with speed $\beta_B$ along a direction (jet axis) inclined with respect to the observer's LOS by an angle $\theta_\LOS$. The corresponding Lorentz boost factor is $\Gamma_B \equiv (1-\beta_B^2)^{-1/2}$. 

For the (lepto-)hadronic models, the energy spectrum of protons in the blob frame fulfills a single power-law distribution \cite{Kardashev1962, 2012.13302}:
\begin{equation}\label{eq:Power_law_Blob}
    \frac{d\Gamma'_p}{dE'_pd\Omega'} = \frac{1}{4\pi}c_p\left(\frac{E'_p}{m_p}\right)^{-\alpha_p}
\end{equation}
with $\gamma'_{\min,\,p} \le E'_p/m_p\le \gamma'_{\max,\,p}$.
The normalization constant $c_p$ can be computed from the proton luminosity $L_p$ \cite{PhysRevD.82.083514}. 
The proton spectrum in the observer's rest frame can then be rewritten as (see Supplemental Material for details)
\begin{equation}\label{eq:CRSpectrum}
\begin{split}
    \frac{d\Gamma_p}{dT_pd\Omega} 
    =&\, \frac{1}{4\pi}c_p\,\left(1+\frac{T_p}{m_p}\right)^{-\alpha_p}\\ &\times\frac{\beta_p(1-\beta_p\beta_B  \mu)^{-\alpha_p} \Gamma_B^{-\alpha_p}}{\sqrt{(1-\beta_p \beta_B \mu)^2 - (1-\beta_p^2)(1-\beta_B^2)}}\,,
    \end{split}
\end{equation}
where $T_p\equiv E_p - m_p$ is the proton kinetic energy, $m_p\simeq 0.938$ GeV is the proton mass, $\beta_p = \left[1-m_p^2/(T_p+m_p)^2\right]^{1/2}$ is the proton speed. Given a SED, the minimal and maximal Lorentz boost factors, i.e. $\gamma'_{\min,\,p}$ and $\gamma'_{\max,\,p}$, the power-law index $\alpha_p$, the Doppler factor $\doppler = [\Gamma_B\left(1-\beta_B\cos\theta_\LOS\right)]^{-1}$, and the luminosity $L_p$ are fitted. Two common assumptions in the fit are $\doppler = 2\Gamma_B$ and $\Gamma_B$, corresponding to, respectively, $\theta_\LOS = 0$ and $1/\doppler$ (with $\doppler\gg1$). We use the results presented in Refs.~\cite{TXS, TXS_2} for TXS 0506+056 and \cite{1304.0605} BL Lacertae, summarized in Table \ref{Tab:HadronicModel}. Additionally, the redshift $z$ \cite{BLRedshift, 1802.01939}, the luminosity distance $d_L$, and BH mass $M_\BH$ (in units of solar masses $M_\odot$) \cite{Titarchuk:2017jwu, 1901.06998} are also given.\\

\begin{table}
\centering
\begin{ruledtabular}
\begin{tabular}{ccc}
    \rowcolor[gray]{.95}
    \multicolumn{3}{@{}c@{}}{\bf (Lepto-)Hadronic Model Parameters}\\
    \colrule
    ~~~Parameter (unit)~~~ & TXS 0506+056~~~ & BL Lacertae~~~ \\
    \colrule
     $z$   &0.337 & 0.069 \\
     $d_L (\text{Mpc})$    & 1835.4 & 322.7 \\
     $M_\BH$ ($M_\odot$) &$3.09\times10^{8} $ &$8.65\times 10^7$\\
    $\mathcal{D}$  &40$^\star$& 15  \\
    $\Gamma_B$ &20& 15  \\
     $\theta_\LOS (^\circ)$  &$0$&$3.82$  \\
    $\alpha_p$   & $2.0$& $2.4$\\
    $\gamma'_{\min,\,p}$ & 1.0 & 1.0  \\
     $\gamma'_{\max,\,p}$ & $5.5\times10^{7^\star}$& $1.9\times10^9$  \\
     $L_p$ (erg/s)  & $2.55 \times 10^{48^\star}$& $9.8 \times 10^{48}$  \\
    \bottomrule
\end{tabular}
\end{ruledtabular}
\caption{The model parameters for the blazars TXS 0506+056 (leptohadronic) \cite{TXS, TXS_2} and BL Lacertae (Hadronic) \cite{1304.0605} used in our calculations. 
The quantities flagged with a star ($^\star$) correspond to mean values computed from the ranges given in the second column of Table 1 of Ref.~\cite{TXS_2} (more details on the impacts of these parameters on the final results are given in Supplemental Material). In the model fitting, the assumption of $\doppler = 2\Gamma_B$ ($\Gamma_B$) is used for TXS 0506+056 (BL Lacertae).}
\label{Tab:HadronicModel}
\end{table}

\begin{figure}
\centering
\includegraphics[width=0.476\textwidth]{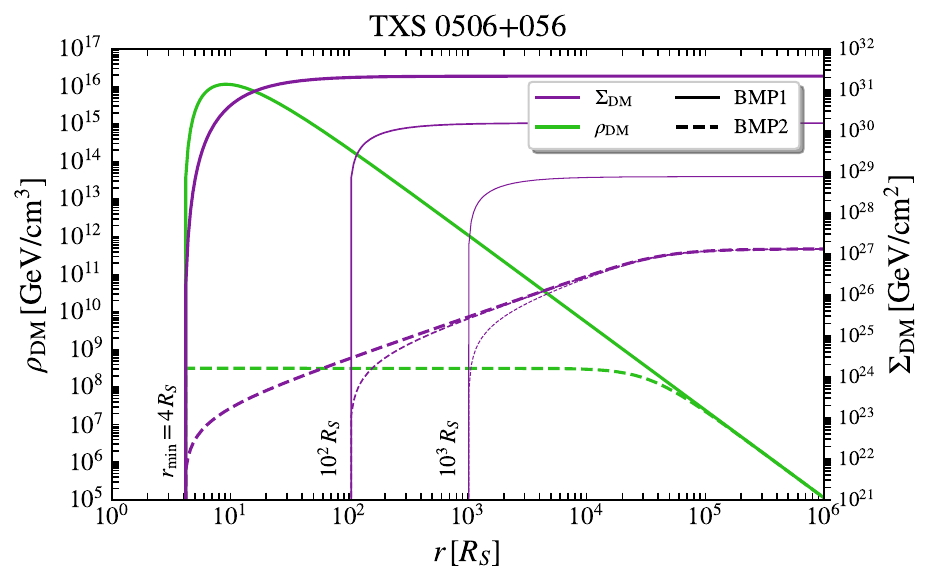}
\caption{The distribution of $\Sigma_\DM$ (purple) and $\rho_\text{DM}$ (green) for TXS 0506+056 with $m_\chi=1$ MeV. The solid and dashed styles correspond to BMP1 and BMP2, respectively. The purple curves, from left to right, are obtained for $r_\min = 4,\,10^2,\,10^3\,R_S$.}\label{fig:sigmaDM}
\end{figure}

\noindent \textit{\textbf{Dark matter density profile.}}--- The adiabatic growth of a BH in the central region of a DM halo is expected to focus the distribution of DM particles, giving rise to a
very dense spike. The phenomenon was first discussed by Gondolo and Silk~\cite{PhysRevLett.83.1719}, who used adiabatic invariants to show
that a preexistent self-gravitating spherical DM profile, with power-law scaling $\rho(r) \propto r^{-\gamma}$, close to the BH is modified into the steeper
profile: 
$
\rho'(r)  \propto  r^{-(9-2\gamma)/(4-\gamma)}\,.
$
While the normalization and radial extension for the spike can be explicitly derived in terms of the normalization of the profile before the BH growth
and the BH mass $M_\BH$, in general one finds that the amount of DM which is displaced to form the spike is about the same as $M_\BH$, see
also Ref.~\cite{Ullio:2001fb}. In the following we  will consider $\gamma=1$ (matching the central scaling of the Navarro-Frenk-White profile, motivated by $N$-body simulations in cold DM cosmologies) 
and fix the normalization via
\begin{equation}\label{eq:DM_condition}
\int_{4R_S}^{10^5 R_S} 4\pi r^2 \rho'(r) dr \simeq M_\BH\,.
\end{equation}
In this expression $R_S$ is the Schwarzschild radius, and the integral extends from $4 R_S$, the radius at which the DM profile goes to zero because of capture onto the BH, to $10^5 R_S$, a typical size for the adiabatically contracted spike. In frameworks with DM candidates that can annihilate in pairs, such as, e.g., thermal relics from the early Universe, there is a maximal DM density compatible with annihilations, about $\rho_\text{core} \simeq m_\chi/(\left\langle \sigma v \right \rangle_0 t_\BH)$, where $\left\langle \sigma v \right \rangle_0$ is the DM annihilation cross section times relative velocity and $t_\BH$ is the time since the BH formed. This may then induce a inner \virg{flattening} of the profile:
\begin{equation}
    \rho_\text{DM}(r) = \frac{\rho'(r) \rho_\text{core}}{\rho'(r) + \rho_\text{core}}.
\end{equation}
Avoiding focusing on specific models, in the following we will refer to two benchmark points (BMPs):
\begin{enumerate}[BMP1)]
    \item $\left\langle \sigma v \right \rangle_0 = 0$, so that $\rho_\text{core}\to +\infty$ and $\rho_\DM = \rho'$;
    \item  $\left\langle \sigma v \right \rangle_0 = 10^{-28} \,\text{cm}^3 \,\text{s}^{-1}$ and $t_\BH = 10^9$ yr;
\end{enumerate}
where the case with $\left\langle \sigma v \right \rangle_0=0$ would be appropriate, e.g., for asymmetric DM models. 
The corresponding profiles for TXS 0506+056 are shown in Fig.~\ref{fig:sigmaDM} together with the LOS integral:
\begin{equation}\label{eq:deltaDM}
    \Sigma_\DM(r) \equiv \int_{r_\min}^{r}  \rho_\DM(r')\,dr'\,,
\end{equation}
where $r_\min$ is the position from where the jet starts. The quantity $\Sigma_\DM$ is relevant for the BBDM signal and tends to saturate at $r\gtrsim 10^5 \,R_S$; a different choice of $\gamma$ or the upper limit of integration $10^5 R_S$ would have a marginal impact. The results on $\Sigma_\DM(r\gtrsim 10^5\,R_S)$ vary with $r_\min$ in the case of BMP1 parameters, while remaining basically invariant for BMP2 if $r_\min\lesssim 10^4\,R_S$. The size of a blazar emitting region is SED model and blazar dependent. In our case $r_\min$ lies within $\sim 10^2\,R_S$ \cite{1304.0605, TXS, TXS_2, PhysRevD.82.083514}, while for other blazars more extremes values up to $10^3 \sim 10^4\,R_S$ are not ruled out \cite{abdo_501,Abdo_2011}. In the further analysis, we will simply adopt the value $r_\min = 4\,R_S$, noting that a different choice below $10^4\,R_S$ would practically correspond to an intermediate situation between the two considered BMPs (see Fig.~\ref{fig:sigmaDM}). The case of BL Lacertae is qualitatively similar.\\

 \noindent \textit{\textbf{Dark matter flux from blazars.}}--- The DM particles can be boosted up to high energies due to elastic scatterings with relativistic protons in the jet.
 Assuming an isotropic scattering and DM at rest, the BBDM flux per kinetic energy reads
\begin{equation}\label{eq:spectrumDM}
    \frac{d\Phi_\chi}{dT_\chi} =\frac{\Sigma_\DM^\text{tot}\,\widetilde{\sigma}_{\chi p}}{2\pi m_\chi d_L^2}\int_{0}^{2\pi}\,d\phi_s\int_{T_p^\min(T_\chi)}^{T_p^\max}\frac{dT_p}{T_\chi^\max(T_p)}\frac{d\Gamma_p}{dT_p d\Omega}\,,
\end{equation}
where $\phi_s$ is the azimuth with respect to the LOS, $T_\chi^\max$ the maximal DM energy after scattering and $\Sigma_\DM^\text{tot}\equiv \Sigma_\DM(r\gg 10^5\,R_S)$; also, we assume
\begin{equation}\label{eq:tildesigma}
    \widetilde{\sigma}_{\chi p} = \sigma_{\chi p} G^2(2m_\chi T_\chi/\Lambda_p^2)\,,
\end{equation}
where $\sigma_{\chi p}\in\{\sigma_{\chi p}^\text{SI},\,\sigma_{\chi p}^\text{SD}\}$ is the zero-momentum transfer spin-independent or spin-dependent cross section and the form factor $G(x^2)\equiv 1/(1 + x^2)^2$ accounts for the proton's internal structure, $\Lambda_p\simeq 0.77$ GeV \cite{PhysRevLett.122.171801}.
The lower extreme of integration $T_p^\min(T_\chi)$ is the minimal kinetic energy the proton should have to pass a kinetic energy $T_\chi$ to DM.
The integral over $T_p$ in Eq.~\eqref{eq:spectrumDM} shows little dependence on the upper extreme of integration because the proton spectrum is attenuated at large energies. We find that, for the purpose of our numerical calculations, fixing $T_p^\max = 10^8$ GeV is accurate enough. We refer to Supplemental Material for more kinematical details.\\

\noindent \textit{\textbf{Direct detection constraints.}}--- BBDM possesses enough energy to leave a signal at direct DM detectors (e.g. XENON1T \cite{XENON:2017lvq})
as well as neutrino detectors (e.g., MiniBooNE \cite{MiniBooNE:2008paa} and Borexino \cite{Borexino:2000uvj}). 
From the top of the atmosphere to the location of the detector, the flux of BBDM will be attenuated due to the scatterings with nucleus $N$ in the air and/or soil \cite{Starkman:1990nj,Mack:2007xj,Hooper:2018bfw,Emken:2018run}.
After having traveled a distance $x$ in the medium, the DM particle remains with a kinetic energy \cite{Emken:2018run,PhysRevLett.122.171801}
\begin{equation}
    T_{\chi}(x)=\frac{2m_{\chi} T_{\chi}e^{-x/\ell}}{2m_{\chi} + T_{\chi}  - T_{\chi} e^{-x/\ell}}\,,
    \label{eq:Txz}
\end{equation}
where $
    \ell^{-1} = \sum_N 2 m_N m_\chi n_N \sigma_{\chi N} /(m_N+m_{\chi})^2
$
is the DM inverse mean free path, with $n_N$ and $m_N$ being the number density and mass of nucleus $N$ in the medium, $\sigma_{\chi N}$ the DM-nucleus cross section. Intuitively, the larger $\sigma_{\chi p}$ is, the more the DM flux is reduced, leading to a blind spot for direct DM detection if $ T_{\chi}(x)$ becomes smaller than the detector's energy threshold $T_\text{exp}^\text{min}$. 
By inverting Eq.~\eqref{eq:Txz},
we approximate the upper limit for $\sigma_{\chi p}$ (dubbed $\sigma_{\chi p}^\text{upper}$) as:
\begin{equation}
    \sigma_{\chi p}^\text{upper} \simeq \log\left[1+\frac{2 m_\chi}{T_\chi^\text{min}(T_\text{exp}^\text{min})}\right] \frac{\sigma_{\chi p}\,\ell}{x}\,,
    \label{eq:upper}
\end{equation}
where $T_\chi^\min(T_\text{exp}^\min)$ is the minimal DM kinetic energy necessary to leave a detectable recoil energy at the direct detector.
Note that $\sigma_{\chi p}\ell$ is actually independent of the cross section. A complication arises in the calculation of $\ell$. In Ref.~\cite{PhysRevLett.122.171801}, the authors use DarkSUSY \cite{Bringmann:2018lay} to calculate the average density $n_N$ of Earth's 11 most abundant elements between the surface and depth $x$. In our work we adopt a more concise and practical approach by using the concept of meter water equivalent (MWE). More specifically, we consider the medium as just composed by water and convert the detector depths in MWE, which for XENON1T, MiniBooNE and Borexino result in 3650~\cite{Harnik:2020ugb}, 26~\cite{MiniBooNE:2008paa} and 3800 MWE~\cite{Borexino:2000uvj}, respectively. The results of our simplified method are in good agreement with those presented in Ref.~\cite{PhysRevLett.122.171801}. The depth $x$ in the Eq.~\eqref{eq:upper} should include the time-dependence effects of the blazar's position with respect to the detector, but we have verified that, for the two considered sources, these would only slightly affect our final results. Moreover, these effects could eventually be avoided by averaging over the full set of blazars in the entire sky.

Whereas, if $\sigma_{\chi p}$ is too small, the BBDM flux and the DM-proton scattering is too weak to leave any recoil in the detectors. Correspondingly, there exists a lower detectable bound on $\sigma_{\chi p}$ which is determined by the detector's sensitivity. Considering an elastic scattering between DM and the target nucleus $N$ and denoting with $T_N$ the nuclear recoil energy, the BBDM induced target nucleus recoil rate can be expressed as
\begin{equation}\label{eq:DM-N rate}
    \Gamma_N^\text{DM} =  \int_{T_\text{exp}^\text{min}}^{T_\text{exp}^\text{max}}dT_N\,\widetilde{\sigma}_{\chi N} \int_{T_{\chi}^\min(T_N)}^{+\infty}\!\!\ 
  \frac{ dT_{\chi}}{T_{N}^{\mathrm{ max}}(T_\chi)}\frac{d\Phi_{\chi}}{dT_{\chi}}\,, 
\end{equation}
where $\left[T_\text{exp}^\min,~T_\text{exp}^\max\right]$ is the energy range of sensitivity of the detector and $T_N^\max$ is the maximal recoil energy of the nucleus. The nuclear cross section $\widetilde{\sigma}_N$ contains the form factor as in Eq.~\eqref{eq:tildesigma}. We emphasize that, since $\widetilde{\sigma}_{\chi N}\propto \sigma_{\chi p}$ and $d\Phi_\chi/dT_\chi \propto \sigma_{\chi p}$, then $\Gamma_N^\DM\propto \sigma_{\chi p}^2$.
By comparison with the nucleus recoil limits of different experiments, we can derive the bounds on $\sigma_{\chi p}$. 

For the spin-independent case, we consider the experiments XENON1T and MiniBooNE. The target nucleus of XENON1T is Xe ($\Lambda_\text{Xe} \approx 141$ MeV \cite{ANGELI2004185}) and the limiting scattering rate per nucleus is given by $
    \Gamma_N (4.9~\text{keV} \leq T_\text{Xe} \leq 40.9~\text{keV} ) < 2.41 \times 10^{-34} ~\text{s}^{-1} 
$
For the MiniBooNE experiment, the limiting counting rate per proton is 
$
    \Gamma_\text{p} (T_p > 35 ~\text{MeV}) < 1.5 \times 10^{-32} ~\text{s}^{-1}
$ \cite{PhysRevLett.122.171801}.
\begin{figure}
\centering
\includegraphics[width=0.45\textwidth]{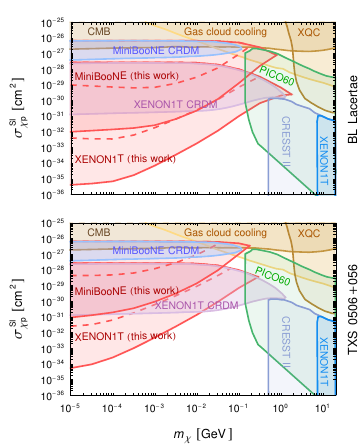}
\caption{The constraints on spin-independent DM-proton cross section imposed by XENON1T \cite{XENON:2017lvq} and MiniBooNE \cite{MiniBooNE:2008paa}. The solid and dashed red lines correspond to BMP1 and BMP2, respectively. The top (bottom) panel is for BL Lacertae (TXS 0506+056). For comparison, the constraints from CRDM \cite{PhysRevLett.122.171801}, cosmic microwave background
(CMB) observations \cite{Xu:2018efh}, gas cloud cooling \cite{Bhoonah:2018wmw}, the x-ray
quantum calorimeter experiment (XQC) \cite{Mahdawi:2018euy}, and a selection of
direct detection experiments \cite{XENON:2017lvq,CRESST:2017ues, CRESST:2015txj, PICO:2017tgi} are included.}
\label{fig:SPB_SI_proton}
\end{figure}
\begin{figure}
\centering
\includegraphics[width=0.43\textwidth]{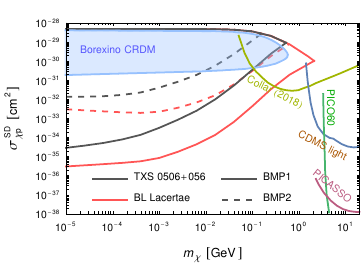}
\caption{The constraints on spin-dependent DM-proton cross section imposed by Borexino \cite{Borexino:2000uvj}. The red (black) lines refer to BL Lacertae (TXS 0506+056) with the solid and dashed styles corresponding to BMP1 and BMP2, respectively. For comparison, the limits from CDMS light \cite{SuperCDMS:2017nns}, PICO60 \cite{PICO:2017tgi}, PICASSO \cite{Behnke:2016lsk}, and Collar \cite{Collar:2018ydf} are also reported.}
\label{fig:SPB_SD_proton}
\end{figure}
The resulting limits on the spin-independent cross section $\sigma_{\chi p}^\text{SI}$ from TXS 0506+056 and BL Lacertae are shown in the upper and lower panels of Fig.~\ref{fig:SPB_SI_proton}, respectively. 
The solid (dashed) lines correspond to BMP1 (BMP2). For each blazar, the difference between solid and dashed lines comes from $\Sigma_\DM^\text{tot}$, and $\sigma_{\chi p} \propto 1/\sqrt{\Sigma_\DM^\text{tot}}$. The sensitivity of BBDM is orders of magnitude higher than that of cosmic ray dark matter (CRDM) \cite{PhysRevLett.122.171801}. Other complementary limits are also shown for comparison.

For the spin-dependent case, the limiting scattering rate per proton can be derived from proton up-scattering in neutrino detectors like Borexino \cite{Borexino:2000uvj}, that is
$
    \Gamma_\text{p} (T_p > 25 ~\text{MeV}) < 2 \times 10^{-39} ~\text{s}^{-1}\,,
$
where we have used the approximation that the ratio between quenched energy deposit (equivalent electron energy $T_e$) and proton recoil energy $T_p$ fulfills $T_e(T_p)/T_p \approx 2$ for $T_p \gtrsim 5$ MeV \cite{Beacom:2002hs,Dasgupta:2011wg}.
We show the constraints on the spin-dependent cross section $\sigma_{\chi p}^\text{SD}$ in Fig.~\ref{fig:SPB_SD_proton}. Again, the sensitivities from BBDM are much stronger than that from CRDM.\\

\noindent \textit{\textbf{Conclusion.}}--- Because of extremely powerful jets and large DM densities, we find that blazars are ideal DM boosters and can induce a DM flux at Earth stronger than the analogous flux due to the boosting of DM particles in the Milky Way halo by galactic cosmic rays. We have focused on two sample sources, TXS 0506+056 — tentatively identified as a high-energy neutrino source — and the closer BL Lacertae. The limits we have derived from the null detection of the connected DM recoil signal (with DM and neutrino detectors) are the most stringent constraints to date on the DM-proton scattering cross section $\sigma_{\chi p}$ for DM masses lighter than about 1~GeV, considering both spin-independent and spin-dependent interactions. The improvement compared to previous results can be as large as 1 up to 5 orders of magnitude, depending on the source and the related uncertainties.

We remark that the results presented here, driven by IceCube observations,
are based on (lepto-)hadronic SED models. For purely leptonic frameworks the situation would be different since protons are much less energetic ($\gamma'_{\max,\, p} \simeq 1$) and their luminosity is in general smaller by several orders of magnitude ($L_p\sim 10^{44}$ erg/s). Naively, we estimate that in these scenarios the BBDM flux and corresponding $\sigma_{\chi p}$ constraints would be far weaker, but precise calculations in such frameworks lie outside the scope of this work.

While the results presented here rely on assumptions regarding the model of individual blazars and of the associated DM density, we expect that extending the analysis to a full blazar ensemble would eventually allow us to significantly reduce the dependence on modelization uncertainties and possibly enhance our results. Besides, while we are suggesting here a novel method to investigate DM properties, this work could also be relevant to improve the current understanding of blazar jet characteristics.\\ 

\begin{acknowledgments}
The authors are grateful to Serguey T. Petcov for useful discussions and helpful suggestions. This work was supported by the research grant \virg{The Dark Universe: A Synergic Multi-messenger Approach} No. 2017X7X85K under the program PRIN 2017 funded by the The Italian Ministry of Education, University and Research (MIUR), and by the European Union’s Horizon 2020 research and innovation program under the Marie Skłodowska-Curie Grant Agreement No. 860881-HIDDeN.
\end{acknowledgments}


%

\newpage
\onecolumngrid
\fontsize{12pt}{14pt}\selectfont
\setlength{\parindent}{15pt}
\setlength{\parskip}{1em}
\newpage
\begin{center}
	\textbf{\large Direct Detection Constraints on Blazar-Boosted Dark Matter} \\ 
	\vspace{0.05in}
	{ \it \large Supplemental Material}\\ 
	\vspace{0.05in}
	{Jin-Wei Wang, Alessandro Granelli and Piero Ullio}
	\vspace{0.05in}
\end{center}
\centerline{{\it  Scuola Internazionale Superiore di Studi Avanzati (SISSA), via Bonomea 265, 34136 Trieste, Italy}}
\centerline{{\it  INFN, Sezione di Trieste, via Valerio 2, 34127 Trieste, Italy}}
\centerline{{\it Institute for Fundamental Physics of the Universe (IFPU), via Beirut 2, 34151 Trieste, Italy}}
\vspace{0.05in}
\setcounter{page}{1}

In this Supplemental Material, we first describe how to derive the jet spectrum in the observer's frame. We then report the kinematical formulae for a generic elastic scattering and present a more detailed derivation of the Blazar-Boosted Dark Matter (BBDM) flux, as given in Eq.~\eqref{eq:spectrumDM} of the main text. Finally, we discuss how variations of the SED model parameters would affect our final results on $\sigma_{\chi p}$ constraints.

\noindent \textit{\textbf{Jet spectrum in the observer's frame}}-- Consider a jet particle with mass $m$ and energy $E$ in the observer's frame, moving in the direction of polar angle $\theta$ and azimuth $\phi$ with respect to the jet axis. The boost factor ($\gamma = E/m$) and polar angle in the blob frame can be derived from a Lorentz transformation:
\begin{eqnarray}
 \gamma'(\gamma,\mu) &=& (1-\beta_B \beta \mu)\gamma \Gamma_B\,,\\ \mu'(\gamma,\mu) &=& \frac{\beta \mu - \beta_B}{\sqrt{(1-\beta_B \beta \mu)^2 - (1-\beta^2)(1-\beta_B^2)}}\,,
\end{eqnarray}
where $\beta = \sqrt{1-1/\gamma^2}$ is the velocity of the particle. 
In our notation the variables with and without a prime are computed in the blob and observer's frame, respectively.
We define the number of particles injected by the blazar per unit time, per unit energy and per unit solid angle as $d\Gamma/(dEd\Omega)$ (dubbed spectrum). 
The spectrum in the observer's frame can be obtained by boosting it from the blob frame. Following similar steps as the ones presented in Ref.~\cite{PhysRevD.82.083514}, we arrive to
\begin{equation}
\begin{split}
    \frac{d\Gamma}{dEd\Omega} 
    =&\, \Gamma_B\frac{d\Gamma'}{dE'd\Omega'}
    \left|\text{det}\begin{pmatrix}
    \frac{\partial \gamma'}{\partial \gamma} &  \frac{\partial \gamma'}{\partial \mu}\\
    \frac{\partial \mu'}{\partial \gamma} &
    \frac{\partial \mu'}{\partial \mu}
    \end{pmatrix}
    \right|\\
    =&\, \frac{d\Gamma'}{dE'd\Omega'} \frac{\beta}{\sqrt{(1-\beta \beta_B \mu)^2 - (1-\beta^2)(1-\beta_B^2)}}\,.
    \end{split}
\end{equation}
It is straightforward to verify that, for a spectrum in the blob frame, e.g.~a single power-law distribution, the spectrum of protons can be reduced to the form given in Eq.~\eqref{eq:CRSpectrum} of the main text.
\begin{figure}
\centering
\includegraphics[width=0.45\textwidth]{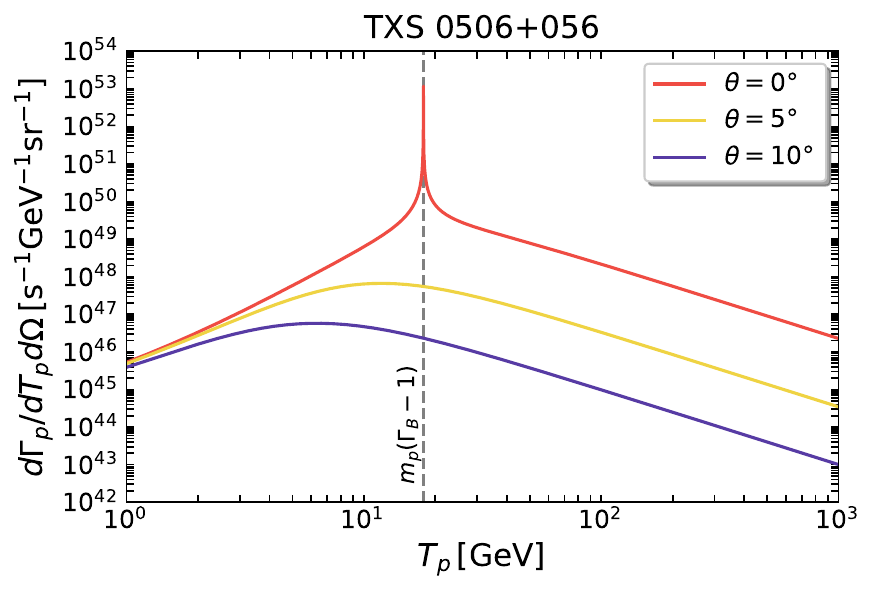}
\includegraphics[width=0.45\textwidth]{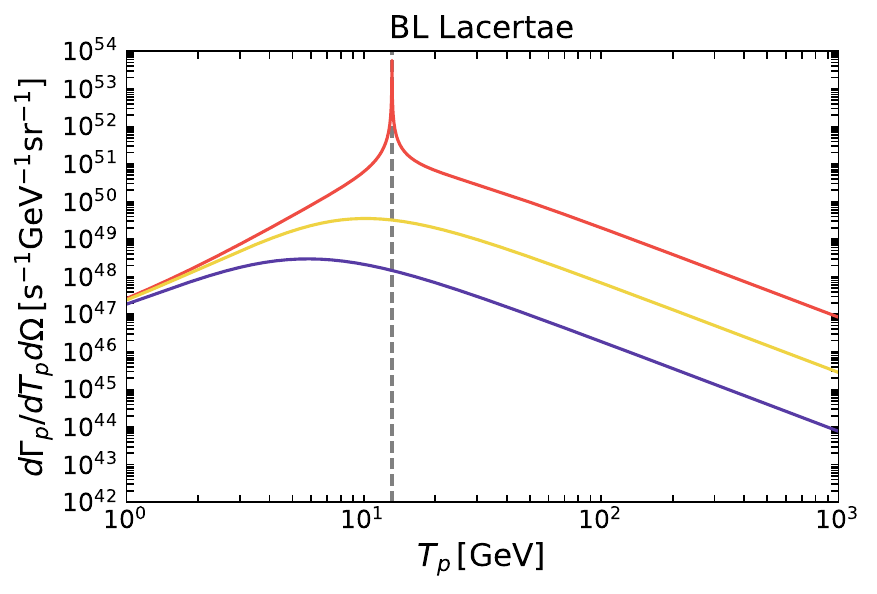}
\caption{\fontsize{12pt}{14pt}\selectfont The spectrum of protons in the observer's frame for TXS 0506+056 (left panel) and BL Lacertae (right panel). The model parameters used are given in Table \ref{Tab:HadronicModel} of the main text. The different colours correspond to different polar angles: $\theta = 0^\circ$ (red), $5^\circ$ (yellow), and $10^\circ$ (purple). The vertical grey dashed lines correspond to proton kinetic energies of $T_p = m_p(\Gamma_B -1)$.}\label{fig:CRSpectrum}
\end{figure}
In Fig.~\ref{fig:CRSpectrum} we show the spectrum of protons from the sources TXS 0506+056 (left) and BL Lacertae (right), for different polar angles and the parameters given in Table~\ref{Tab:HadronicModel} of the main text.
For the high-energy region ($T_p \gg m_p$), the spectra are parallel because $d\Gamma_p/(dT_pd\Omega)\propto T_p^{-\alpha_p}$ (see Eq.~\eqref{eq:CRSpectrum}), while for the low-energy region ($T_p \ll m_p$), we have $d\Gamma_p/(dT_pd\Omega) \approx c_p \Gamma_B^{-\alpha_p}(4\pi \beta_B)^{-1} \sqrt{2T_p/m_p}$. 
The peak that appears in the curve for $\theta = 0^\circ$ corresponds to the kinetic energy of a proton that is at rest in the blob frame, i.e. $T_p = m_p(\Gamma_B -1)$.\\

\noindent \textit{\textbf{Elastic scattering kinematics}}-- For an elastic scattering process $i,\,j \to i,\,j$, where $i,\,j$ denote particles with mass $m_{i,\,j}$. In the main text we consider the case in which $i,\,j \in \{\chi,\,p,\,N\}$. In the laboratory (LAB) frame, we assume that $j$ is effectively at rest. After a scattering, the kinetic energy (or recoil energy) $T_j$ transferred to the particle $j$ from the incoming particle $i$ with kinetic energy $T_i$ is \cite{PhysRevLett.122.171801}
\begin{equation}
    T_j = T_j^\max\, \frac{1+\mu_s^*}{2}~~~\text{with}~~ T_j^\max(T_i) =  \frac{\left(T_i^2 + 2m_i T_i\right)}{T_i + (m_i+m_j)^2/(2m_j)},
    \label{eq:Tjmax}
\end{equation}
where $\mu_s^*$ is the cosine of the scattering angle in the center-of-mass (c.m.) rest frame.
Inverting Eq.~\eqref{eq:Tjmax} gives the minimal energy the particle $i$ should have to pass a kinetic energy $T_j$ to particle $j$ \cite{PhysRevLett.122.171801}:
\begin{equation}
    T_i^\min(T_j) = \left(\frac{T_j}{2}-m_i\right)\left[1\pm\sqrt{1+ \frac{\left(m_i+m_j\right)^2}{\left(T_j-2m_i\right)^2} \frac{2T_j}{m_j}}\right],
    \label{eq:Tpmin}
\end{equation}
where the $+ (-)$ applies for $T_j \geq 2m_i$ ($T_j<2m_i$). Lorentz transformations relate the scattering angles in the LAB and c.m.~frames via:
\begin{equation}\label{eq:mustar}
    \mu^*_s=\frac{2\mu_s^2}{\mu_s^2+\gamma_\CM^2(T_i)(1-\mu_s^2)}-1\,,
\end{equation}
where
\begin{equation}
\gamma_\CM^2(T_i) \equiv \frac{(T_i+m_i+m_j)^2}{\left(m_i+m_j\right)^2+2m_j T_i}\,.
\end{equation}
Using Eq.~\eqref{eq:mustar} and assuming an isotropic collision in the c.m.~frame, we can derive the probability distribution of the scattering angle for the particle $j$ in the LAB frame:
\begin{equation}
    P(\mu_s;T_i)\equiv \frac{1}{2} \,\frac{d\mu_s^*}{d\mu_s} =\frac{2\mu_s\gamma_\CM^2(T_i)\Theta(1-\mu_s)}{\left[\mu_s^2+\gamma_\CM^2(T_i)(1-\mu_s^2)\right]^2}\,.
\end{equation}
where the Heaviside theta function ensures that $0\leq\mu_s\leq1$. The kinetic energy of the outgoing particle $j$ in terms of the scattering angle in the LAB frame is then given by:
\begin{equation}\label{eq:Tj(mus)}
    T_j(T_i,\mu_s) =  T^\max_j(T_i) \frac{\mu_s^2}{\mu_s^2+\gamma^2_\CM(T_i)(1-\mu_s^2)}.
\end{equation}
Inverting Eq.~\eqref{eq:Tj(mus)}, we obtain the scattering angle that corresponds to an incoming kinetic energy $T_i$ and transferred energy $T_j$:
\begin{equation}
    \overline{\mu}_s(T_i, T_j) = \left[1 + \frac{T_j^\max(T_i)-T_j}{T_j \gamma_\CM^2(T_i)}\right]^{-1/2}.
\end{equation}

\noindent {\bf \textit{BBDM flux at Earth}} -- We now briefly explain how to derive the BBDM flux as given in Eq.~\eqref{eq:spectrumDM} of the main text. We make use of the kinematic formulae presented above using $i = p$ and $j = \chi$ and consider the LAB frame to coincide with the observer's frame, i.e.~we neglect the motion of dark matter with respect to the protons. Assuming the elastic cross section to be isotropic in the c.m.~rest frame, the flux per kinetic energy of BBDM can be written as

\begin{equation}
    \frac{d\Phi_\chi}{dT_\chi} = \frac{\Sigma^\text{tot}_\DM\,\widetilde{\sigma}_{\chi p}}{m_\chi d_L^2}\int_{0}^{2\pi}\,d\phi_s\int_{0}^{1}\,d\mu_s
    \int_{T_p^\min(T_\chi)}^{T_p^\max(T_\chi)}
    \,dT_p\,\frac{d\Gamma_p}{dT_p d\Omega}
    \frac{P(\mu_s; T_p)}{2\pi}\delta\left(T_\chi-T_\chi\left(T_p,\mu_s\right)\right)
\end{equation}
where $\phi_s\in[0,2\pi]$ is the azimuthal angle with respect to the LOS. $T_\chi(T_p, \mu_s)$ is the dark matter kinetic energy after scattering for given $T_p$ and $\mu_s$. Note that the proton spectrum depends on $\mu$, which is related to $\mu_s$ and $\phi_s$ by a rotation of an angle $\theta_\LOS$, namely
\begin{equation}
    \mu(\mu_s,\phi_s) = \mu_s \cos\theta_\LOS + \sin\phi_s\sin\theta_\LOS\sqrt{1-\mu_s^2}\,.
\end{equation}
Using the properties of the Dirac $\delta$-function we can write:
\begin{equation}
    \delta\left(T_\chi-T_\chi(T_p,\mu_s)\right)
    =  \frac{\delta\left(\mu_s-\overline{\mu}_s(T_p, T_\chi)\right)}{T_\chi^\max(T_p)P(\mu_s;T_p)},
\end{equation} 
Therefore, the flux can be rewritten as:
\begin{equation}\label{eq:spectrumDM_2}
    \frac{d\Phi_\chi}{dT_\chi} =\frac{\Sigma^\text{tot}_\DM\,\widetilde{\sigma}_{\chi p}}{2\pi m_\chi d_L^2}\int_{0}^{2\pi}\,d\phi_s\int_{T_p^\min(T_\chi)}^{T_p^\max}\frac{dT_p}{T_\chi^\max(T_p)}\frac{d\Gamma_p}{dT_p d\Omega}\,,
\end{equation}
where the proton spectrum needs to be evaluated at $\mu(\overline{\mu}_s(T_p,T_\chi),\phi_s)$. 
We note that, in the case of $\theta_\LOS = 0$, $\mu = \mu_s$ and therefore the integration over $\phi_s$ in Eq.~\eqref{eq:spectrumDM_2} becomes trivial.
\begin{figure}
\centering
\includegraphics[width=0.45\textwidth]{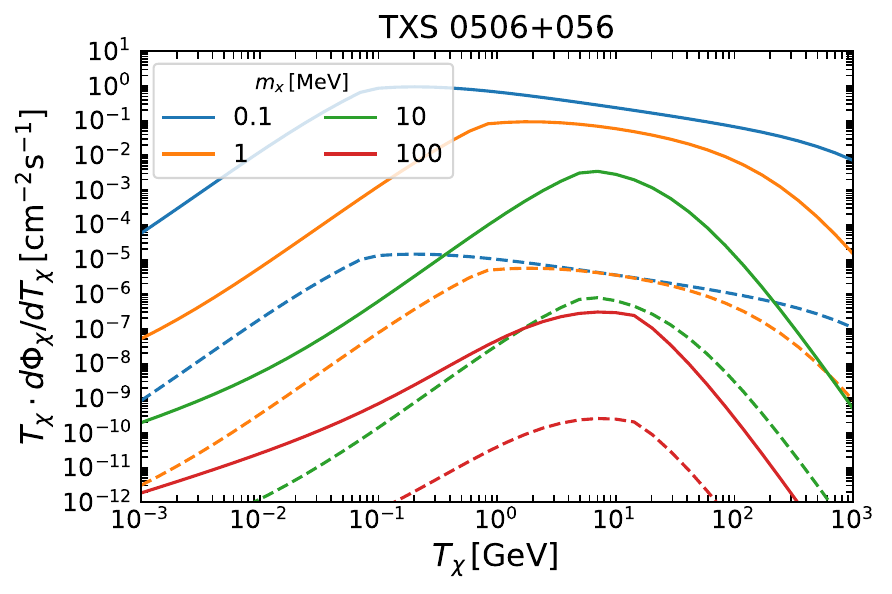}
\includegraphics[width=0.45\textwidth]{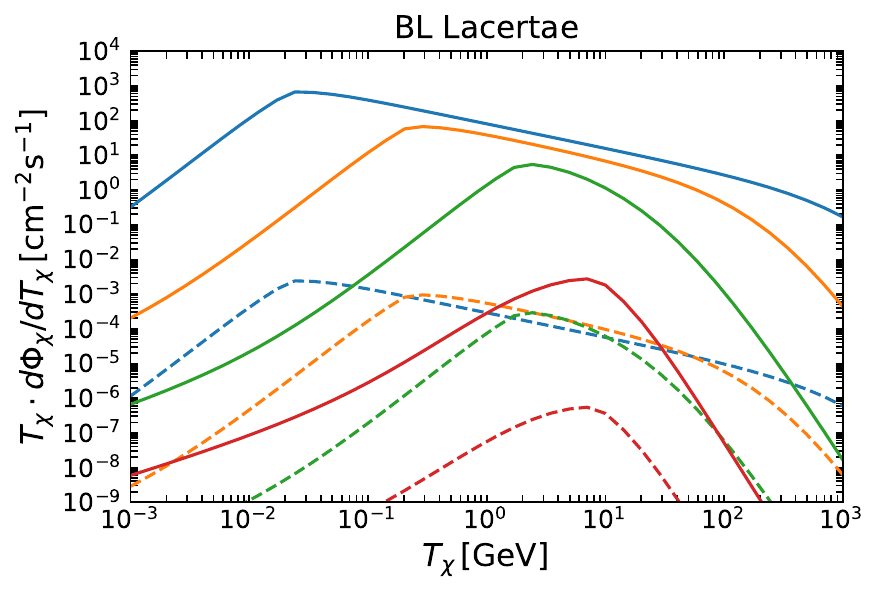}
\caption{\fontsize{12pt}{14pt}\selectfont The expected flux of BBDM from TXS 0506+056 (left panel) and BL Lacertae (right panel). Different colours correspond to different DM mass $m_\chi$, namely 0.1 MeV (blue), 1 MeV (orange), 10 MeV (green), and 100 MeV (red). The solid and dashed lines represent BMP1 and BMP2, respectively. Note that all these results are derived by setting $\sigma_{\chi p} = 10^{-30}~\text{cm}^2$.}
\label{fig:DM_Tx_Spectrum}
\end{figure}

Using the results for $\Sigma_\DM$ discussed in the main text, the parameters in Table \ref{Tab:HadronicModel}
and a typical value $\sigma_{\chi p}=10^{-30}\,\text{cm}^2$, we compute numerically the integrals in Eq.~\eqref{eq:spectrumDM_2} and plot in Fig.~\ref{fig:DM_Tx_Spectrum} the BBDM spectrum for the sources TXS 0506+056 (left) and BL Lacertae (right) for a few DM masses.

\noindent {\bf \textit{The dependence on the SED model parameters}} -- As we have mentioned in the main text, the blazar jet physics can be well formalised by the \virg{blob geometry}, and the relevant model parameters can be derived by fitting the observations of the photon SED. In our analysis, four important parameters are involved, namely $L_p$, $\gamma'_{\max,\,p}$, $\Gamma_p$ and $\alpha_p$ (see Table \ref{Tab:HadronicModel} of the main text), and, for some of them, we have simply adopted the mean values from Refs.~\cite{TXS_2,1304.0605}. 
In order to give a more robust analysis,
here we investigate the effects of different choices of the SED model parameters on our final results.
For brevity, in this section we only consider the blazar TXS 0506+056 and ignore the case of BL Lacertae since the results are qualitatively similar.

The dependence of our final results on $L_p$ is quite straightforward. The proton luminosity only enters in the normalization factor of the proton spectrum as $c_p \propto L_p$ (see Eqs.~\eqref{eq:Power_law_Blob} and \eqref{eq:CRSpectrum}). Then, from Eq.~\eqref{eq:DM-N rate} one can easily obtain that the lower exclusion boundary on $\sigma_{\chi p}$ is proportional to $ 1/\sqrt{L_p}$. Specifically, in Ref.~\cite{TXS_2} $L_p$ ranges from $1.6 \times 10^{48}$ to $3.5 \times 10^{48}$ erg/s, so its influence on the $\sigma_{\chi p}$ lower bound is about 20\%. Concerning $\gamma'_{\max,\,p}$, when it varies from $4\times10^7$ to $7\times10^7$ \cite{TXS_2}, we find that its effects on the $\sigma_{\chi p}$ constraints are negligible ($\lesssim 1\%$). 
This is because $\gamma'_{\max,\,p}$ only affects the high-energy part of the proton spectrum, which is exponentially suppressed.

Instead, the influences of the Lorentz factor $\Gamma_B$ and the proton spectrum power index $\alpha_p$ on the $\sigma_{\chi p}$ bounds are more complicated. In order to show their effects explicitly, in Table \ref{Tab:uncertainties} we present the limitations on the logarithm of the DM-proton cross section ($\text{log}_{10}[\sigma_{\chi p}/\text{cm}^2]$) from the blazar TXS 0506+056 for some different choices of $\alpha_p$ and $\Gamma_B$. Note that the other SED parameters are fixed to the same values in Table \ref{Tab:HadronicModel}, while the quantities related to the DM density profile are as in BMP1. 
The selected values for $\Gamma_B$ are chosen according to the range given in Ref.~\cite{TXS_2}.
In contrast, the value of $\alpha_p$ is a constant in Ref.~\cite{TXS_2}, so few different values are manually selected to examine its influence. 

The effects of $\Gamma_B$ and $\alpha_p$ on the $\sigma_{\chi p}$ constraints are as follows. For larger (smaller) values of $\Gamma_B$, the energy of protons in the BH frame is also larger (smaller) while, conversely, the total number of protons emitted per seconds by the source diminishes (increases). The overall effects on the BBDM flux, and, consequently, on the DM-proton cross section constraints, depend on the DM mass. More specifically, for a larger $\Gamma_B$, the constraints on $\sigma_{\chi p}$ becomes more stringent in the low-mass region ($m_\chi \lesssim 10^{-4}$ GeV) and weaker for higher masses ($m_\chi \gtrsim 10^{-4}$ GeV), while for a smaller $\Gamma_B$ the results are opposite. 
In any case, the effects of varying $\Gamma_B$ in the range $[17.5, ~22.5]$ only amounts to a $\sim 30\%$ correction.
As for $\alpha_p$, if it is increased (decreased) with a fixed $\Gamma_B$, the BBDM flux will be amplified (reduced) and render a stronger (weaker) constraints on $\sigma_{\chi p}$. For the two selected values of $\alpha_p = 2.5$ and $\alpha_p = 1.8$, the effects on $\sigma_{\chi p}$ correspond to roughly a factor of 3. 

In summary, we conclude that the constraints on $\sigma_{\chi p}$ obtained in the main text are representative, and the impact of choosing other parameter values in the ranges given in Ref.~\cite{TXS_2} is relatively small. 

\newcolumntype{C}{@{}>{\centering\arraybackslash}X}
\begin{table}
\fontsize{12pt}{14pt}\selectfont
\centering
\begin{tabularx}{\textwidth}{@{}CCCCCCC@{}}
\hline
\hline
\rowcolor[gray]{.95}
\addlinespace[0.3ex] 
    \multicolumn{7}{@{}c@{}}{\bf Constraints on DM-proton cross section from TXS 0506+056}\\
    \addlinespace[0.3ex] 
    \hline
\addlinespace[0.3ex] 
\multirow{2}{*}{$\alpha_p$} & \multirow{2}{*}{$\Gamma_B$}~ & \multicolumn{5}{c}{$m_\chi$ (GeV)}\\ 
\addlinespace[0.3ex]
\cline{3-7}
\rule{0pt}{3ex} 
                  &                   & $10^{-5}$ & $10^{-4}$ & $10^{-3}$ &  $10^{-2}$ &  $10^{-1}$\\ 
\hline
\rowcolor[gray]{.95}[0pt][0pt]
\addlinespace[0.3ex]
{\bf 2.0}                 &   {\bf 20}    & {\bf $-$34.27} & {\bf $-$33.59} & {\bf $-$32.32} & {\bf $-$30.86} & {\bf $-$29.14}  \\
\addlinespace[0.3ex]
\hline
\addlinespace[0.3ex]
2.0     &   22.5      & $-34.30$ & $-33.57$ & $-32.26$ & $-30.78$ & $-29.04$  \\
\addlinespace[0.3ex]
2.0 &   17.5     & $-34.23$ & $-33.61$ & $-32.39$ & $-30.96$ & $-29.26$  \\
\addlinespace[0.3ex]
2.5 &   20     & $-34.68$ & $-34.03$ & $-32.73$ & $-31.25$ & $-29.41$  \\
\addlinespace[0.3ex]
1.8     &   20    & $-33.81$ & $-33.11$ & $-31.85$ & $-30.41$ & $-28.75$  \\
\hline
\hline
\end{tabularx}
\caption{\fontsize{12pt}{14pt}\selectfont 
The influence of different power index $\alpha_p$ and Lorentz factor $\Gamma_B$ on the constraints of $\text{log}_{10}[\sigma_{\chi p}/\text{cm}^2]$ from the blazar TXS 0506+056 \cite{TXS_2}. For comparison, the benchmark values that we have adopted in the main text are shown in bold in the shaded row.
Note that the quantities involved in the DM density profile are as in BMP1 and the other SED parameters as in Table I.
}
\label{Tab:uncertainties}
\end{table}

\end{document}